\documentclass[11pt,preprint]{aastex}

\newcommand\alpharad{\alpha_{\mathrm{rad}}}

\newcommand\Jy{\;\mbox{Jy}}
\newcommand\sr{\;\mbox{sr}}
\newcommand\MHz{\;\mbox{MHz}}
\newcommand\GHz{\;\mbox{GHz}}
\newcommand\WHz{\;\mbox{W}\,\mbox{Hz}^{-1}}
\newcommand\MpcCubed{\;\mbox{Mpc}^{-3}}
\newcommand\GpcCubed{\;\mbox{Gpc}^{-3}}

\slugcomment{Accepted for publication in the ApJ on November 2, 2007}

\title{MOJAVE: Monitoring of Jets in AGN with VLBA
Experiments. IV. The Parent Luminosity Function of
Radio-Loud Blazars}

\shorttitle{The Parent Luminosity Function of Radio-Loud
Blazars}

\author{M. Cara}
\affil{Department of Physics, Purdue University, 525
Northwestern Avenue, West Lafayette, IN 47907}
\email{mcara@physics.purdue.edu}
\and
\author{M. L. Lister}
\affil{Department of Physics, Purdue University, 525
Northwestern Avenue, West Lafayette, IN 47907}
\email{mlister@physics.purdue.edu}

\begin{document}

\begin{abstract}

We use a complete sample of active galactic nuclei (AGN)
selected on the basis of relativistically beamed 15 GHz radio
flux density (MOJAVE: Monitoring of Jets in AGN with VLBA
Experiments) to derive the parent radio luminosity function
(RLF) of bright radio-selected blazar cores. We use a maximum
likelihood method to fit a beamed RLF to the observed data and
thereby recover the parameters of the intrinsic (unbeamed) RLF.
We analyze two subsamples of the MOJAVE sample: the first
contains only objects of known FR~II class, with a total of 103
sources, and the second subsample adds 24 objects of uncertain
FR class for a total of 127 sources. Both subsamples exclude
four known FR~I radio galaxies and two gigahertz-peaked
spectrum sources. We obtain good fits to both subsamples using
a single power law intrinsic RLF and a pure density evolution
function of the form
$z^m\exp\left[-1/2((z-z_0)/\sigma)^2\right]$. We find that a
previously reported break in the observed MOJAVE RLF actually
arises from using incomplete bins (because of the luminosity
cutoff) across a steep and strongly evolving RLF, and does not
reflect a break in the intrinsic RLF. The derived space density
of the parent population of the FR~II sources from the MOJAVE
sample (with $L_{15\,\mathrm{GHz}}\geq1.3\times 10^{25}\WHz$)
is approximately $1.6 \times 10^3\GpcCubed$.

\end{abstract}


\section{Introduction}

The radio luminosity function (RLF) of active galactic nuclei
(AGN) and its redshift dependence are important quantities in
understanding the physics of AGN and their cosmological
evolution. In the case of AGN selected on the basis of
relativistic emission (i.e., blazars), it can also provide
information about the parent population from which an observed
sample is drawn. A parameterized luminosity function (LF) can
also be useful for producing Monte Carlo simulations of
populations to compare with statistical properties of observed
AGN \citep[e.g.,][]{LM97} as well as to study those properties
of AGN that are difficult to observe directly. The intrinsic
RLF can also be useful for predicting the number of
$\gamma$-ray blazars to be observed by future surveys
\citep[e.g., GLAST; also see][]{LM99} as well as for
determining how rare individual blazars are in the general AGN
population.

According to contemporary AGN unification schemes \citep[see
review by][]{Urry95}, various observed classes of AGN (e.g.,
radio galaxies, quasars, and BL~Lacs) can be the result of
different orientations of essentially the same type of object.
One can test unification schemes using statistical approaches.
For example, if BL~Lac objects are highly beamed versions of
lower power radio galaxies, then the number of BL~Lacs should
be much smaller then the number of parent radio galaxies,
because BL~Lacs are oriented at a small angle to the line of
sight. Previously, \citet{Urry91a}, \citet{Padovani92}, and
\citet{Urry95} applied relativistic beaming corrections (e.g.,
\citealt*{Cohen07}) to the RLF of high power radio galaxies and
found it to be compatible with the observed RLF of a sample of
flat-spectrum, radio-loud quasars. \citet{JW99} proposed a
dual-population unified scheme in which (a) the high-power
FR~II radio galaxies are the parents of all radio quasars and
some BL~Lac-type objects, and (b) moderate-power FR~I radio
galaxies are the parents of the remaining BL~Lac-type objects.
They tested this model by beaming (using Monte Carlo jet
populations with a single bulk Lorentz factor) the
low-frequency radio data and comparing them with high-frequency
radio data.

The MOJAVE AGN sample \citep{LH05} is the first large,
radio-selected AGN sample for which jet kinematic and apparent
superluminal speed information are available
\citetext{\citealp{Kellermann04}; Lister et al. 2008, in
preparation}. It is complete with respect to relativistically
beamed jet emission, and therefore provides a unique
opportunity to learn about the intrinsic (parent) RLF of
blazars. The determination of the intrinsic (non-beamed) RLF is
complicated, however, by relativistic beaming and selection
effects. The radio emission from an AGN is highly enhanced by
Doppler boosting if its jet is relativistic and aligned close
to the line of sight. A flux density-limited sample of AGN will
therefore contain not only sources with high intrinsic
luminosity, but also sources with lower intrinsic luminosities
whose flux densities are Doppler boosted because of their
orientation. The effect of Doppler beaming on the observed RLF
was first calculated for single Lorentz factors \citep{Urry84}
and later extended for distributions of Lorentz factors
\citep{Urry91b}. \citet{L03} extended these studies by deriving
fully analytical expressions for the Doppler factor
distributions and beamed RLFs.

Previous studies \citep[see, e.g.,][]{Padovani92} started with
the assumption that the intrinsic parent LF was that of the
FR~II galaxies, and then applied beaming in order to compare it
with the LFs of the steep spectrum radio quasars (SSRQ) and
flat spectrum radio quasars (FSRQ). Since their sample was
selected at a much lower radio frequency ($2\GHz$), it was
subject to contamination by sources with large extended
emission, as well as uncertainties in the classification of the
sources into SSRQ and FSQR based on a somewhat arbitrary
spectral index cutoff of $\alpharad=-0.5$.  In addition, in
order to simulate the beamed LFs of SSRQ and FSRQ it was
necessary to know the ratio of the core to extended emission.
\citet{Padovani92} assumed a linear relationship between the
beamed and unbeamed luminosities, but found that they needed
different factors to produce good fits to the SSRQ and FSRQ. A
somewhat similar approach was used by \citet{JW99} in
determining the beaming models of the parent populations in
their dual-population unification scheme.

The MOJAVE sample is different from previous samples in that it
is selected on the basis of (highly variable) radio flux
densities at a high frequency of $15\GHz$, thus effectively
eliminating contamination from extended source emission. The
uncertainties surrounding the spectral index cutoff and
core-to-extended emission ratios are likewise alleviated.

In this paper we use the maximum likelihood method to fit a
beamed RLF to the observed data, from which we recover the RLF
parameters of the parent population of the MOJAVE sample. These
parameters will be used in upcoming studies of the effects of
beaming on the blazar properties derived from flux limited
samples.

The outline of the paper is as follows: In \S~\ref{sec:Sample}
we describe the observational sample and our method for dealing
with incomplete redshift information. In \S~\ref{sec:PLF} we
describe our parameterization of the RLF, and in
\S~\ref{sec:MLMethod} we describe the method used to find the
optimized model parameters and constraints on the fits. We
present the results of the model fitting in
\S~\ref{sec:RLFResults} and summarize our findings in
\S~\ref{sec:Conclusions}.

Throughout this paper we assume (unless stated otherwise) a
cosmology with $\Omega_m=0.3$, $\Omega_{\Lambda}=0.7$,
$\Omega_{r}=0$ and $H_0=70\mbox{ km}\mbox{ s}^{-1}\mbox{
Mpc}^{-1}$. All luminosities are quoted as monochromatic
luminosities at specific frequency $\nu$.  We also adopt the
following convention for the spectral index, $\alpharad$:
$S_{\nu}\propto\nu^{\alpharad}$.

\section{Observational Data}\label{sec:Sample}

\subsection{Sample Description}\label{sec:SampleDesc}

The MOJAVE AGN sample \citep{LH05} consists of all 133 known
bright AGN with galactic latitude $|b|>2.5\arcdeg$, J2000.0
declination greater than $-20\arcdeg$, and compact
(VLBA\footnote{The Very Long Baseline Array is operated by the
National Radio Astronomy Observatory, which is a facility of
the National Science Foundation operated under cooperative
agreement by Associated Universities, Inc.}) flux density
exceeding $1.5\Jy$ at 15~GHz ($2\Jy$ for sources with
$\delta<0$) at any epoch between Jan. 1, 1994 -- Dec. 31, 2003.
The sky area covered by the MOJAVE survey is $6.00912\sr$ for
the northern sky and $2.08012\sr$ for the southern sky. The
sample is selected on the basis of beamed jet emission only.
The contribution from the large-scale radio emission is
effectively excluded by using the milliarcsecond scale (VLBA)
15 GHz flux density, since the former tends to be diffuse and
has a steep radio spectrum.

Many of the MOJAVE sources exhibit high flux density
variability and have been selected based on their largest flux
density values, thus potentially creating a source of selection
bias. However, in a detailed study of AGN variability,
\citet{L01} concluded that the effect of the variability on the
sample selection is small in moderately sized samples because
the majority of highly beamed sources in the parent population
\citep[which are preferentially selected in beamed
emission-selected samples; see, e.g.,][]{VC94} would lie well
above the survey flux limit and will be selected regardless of
their flaring levels. Those sources just above the survey limit
would be statistically balanced by other sources lying just
below.

We present basic data for the MOJAVE sample, including
Fanaroff-Riley (FR) and optical classifications for the MOJAVE
sample in Table \ref{tab:Sample}. For optical class we used the
\citet{Veron06} catalog as well as the NASA/IPAC Extragalactic
Database (NED). We note that the optical classification of
blazars into BL~Lacs and optically violently variable (OVV)
quasars remains controversial \citep[see,
e.g.,][]{Antonucci93,Kovalev05} but this is not essential to
our analysis, which is based on Fanaroff-Riley classes. We base
our Fanaroff-Riley classification of the MOJAVE sources on
their extended radio morphology. We assign an uncertain FR
class to BL~Lacs and intermediate BL~Lac/HPQ sources for which
morphological FR classification is difficult (i.e., core, halo,
unusual morphologies or otherwise lack of prominent hotspots).
For quasars only, in situations were morphological
classification is difficult we rely on the source's luminosity
to assign it a FR class. At $178\MHz$ the Fanaroff-Riley divide
occurs at about $L_{178\;\mathrm{MHz}}\approx 10^{25.3}\WHz$,
but this transition depends on the host galaxy magnitude
\citep[see][]{Owen94,Ledlow96} and sources close to this
luminosity can be of either class. The line separating the two
classes at $1.4\GHz$ in Fig. 1 of \citet{Ledlow96} can be
approximated as $\log L_{\mathrm{FR~break}} \approx
10.2-\frac{2}{3}\mbox{M}_{\mathrm{R}}^{\mathrm{host}}$ over the
range $-25<\mbox{M}_{\mathrm{R}}^{\mathrm{host}}<-21$ (assuming
their cosmology of $q_0=0$ and $H_0=75\mbox{ km}\mbox{
s}^{-1}\mbox{ Mpc}^{-1}$). We used this equation to find the
FR~break luminosity, $L_{\mathrm{FR~break}}$, for a given host
magnitude. We considered any quasar with an inconclusive
morphology to be of FR~II class if its luminosity at $1.4\GHz$
is at least an order of magnitude larger than
$L_{\mathrm{FR~break}}$. Unfortunately, host galaxy magnitudes
are available only for six of the $91$ of the quasars in our
sample. \citet{Pagani03} found that the average absolute
magnitude of the host galaxies radio loud quasars at low
redshifts ($z<0.5$) is
$\langle\mbox{M}_{\mathrm{R}}\rangle=-24.0\pm 0.5$, $q_0=0$ and
$H_0=50\mbox{ km}\mbox{ s}^{-1}\mbox{ Mpc}^{-1}$ which when
converted to the cosmology of \citet{Ledlow96} becomes
$\langle\mbox{M}_{\mathrm{R}}\rangle \approx -23.1$. We assume
$\mbox{M}_{\mathrm{R}}=-23.1$ for the $85$ quasars for which we
could not find information in the literature on the absolute
magnitudes of their host galaxies. We have computed the
luminosities at $1.4\GHz$ from the fluxes found on NED (we used
the largest listed flux) assuming a spectral index
$\alpharad=-0.7$. All of the quasars with uncertain
morphological FR classifications ended up as FR~II class
according to this model. Of the eight galaxies in the MOJAVE
sample, three galaxies (0007$+$106: III~Zw~2, 0415$+$379:
3C~111 and 1957$+$405: Cygnus~A) show FR~II morphologies,
another four galaxies (0238$-$084: NGC~1052, 0316$+$413: 3C~84,
0430$+$052: 3C~120 and 1228$+$126: M87) show FR~I morphologies,
and the galaxy 2021$+$614: OW$+$637 is a gigahertz-peaked
spectrum (GPS) source. The other GPS source in the MOJAVE
sample is the quasar 0742$+$103. Redshifts are available from
NED for all but 12 sources (6 optically featureless BL~Lacs,
and 6 sources without optical identifications).

According to a contemporary unification scheme
\citep[e.g.,][]{Urry95}, the parent population of BL~Lacs is
identified with FR~I type galaxies. However, the issue of
parent populations for BL~Lac objects remains under debate.
Recent studies of the host galaxy and extended radio emission
of radio-selected, low energy peaked BL~Lacs
\citep[e.g.,][]{Cassaro99,Rector2001,Kot05} appear to rule out
the FR~I -- BL~Lac unification scheme in its simplest form.

\citet{JW99} proposed a dual-population scheme in which FR~II
radio galaxies are the misaligned parents of flat-spectrum
quasars and some BL~Lacs. We adopt this unification model and
exclude 4 FR~I galaxies (0238$-$084, 0316$+$413, 0430$+$052,
and 1228$+$126) from the sample because they may belong to a
different parent population and exhibit a different evolution
from the rest of the sources. We also exclude the two GPS
sources 0742$+$103 and 2021$+$614. In our analysis we will use
two samples: one containing known FR~II sources (hereafter, the
``known FR~II sample'') and a second sample comprising both
known FR~II and uncertain FR class sources (hereafter, the
``full sample''). The ``known FR~II sample'' contains 103
sources \citep[91 quasars, 3 FR~II galaxies, 1 BL~Lac and 8
sources of the intermediate BL~Lac/High Polarization Quasar
(HPQ) class; e.g.,][]{Veron00}. One of the sources in this
sample lacks redshift information. The ``full sample'' contains
127 sources (91 quasars, 3 FR~II galaxies, 10 BL~Lacs, 17
BL~Lac/HPQs and 6 sources without optical counterparts). In
this sample 12 sources lack redshift information.

\begin{figure}[htb]
\begin{center}
\includegraphics[angle=-90, scale=0.65]{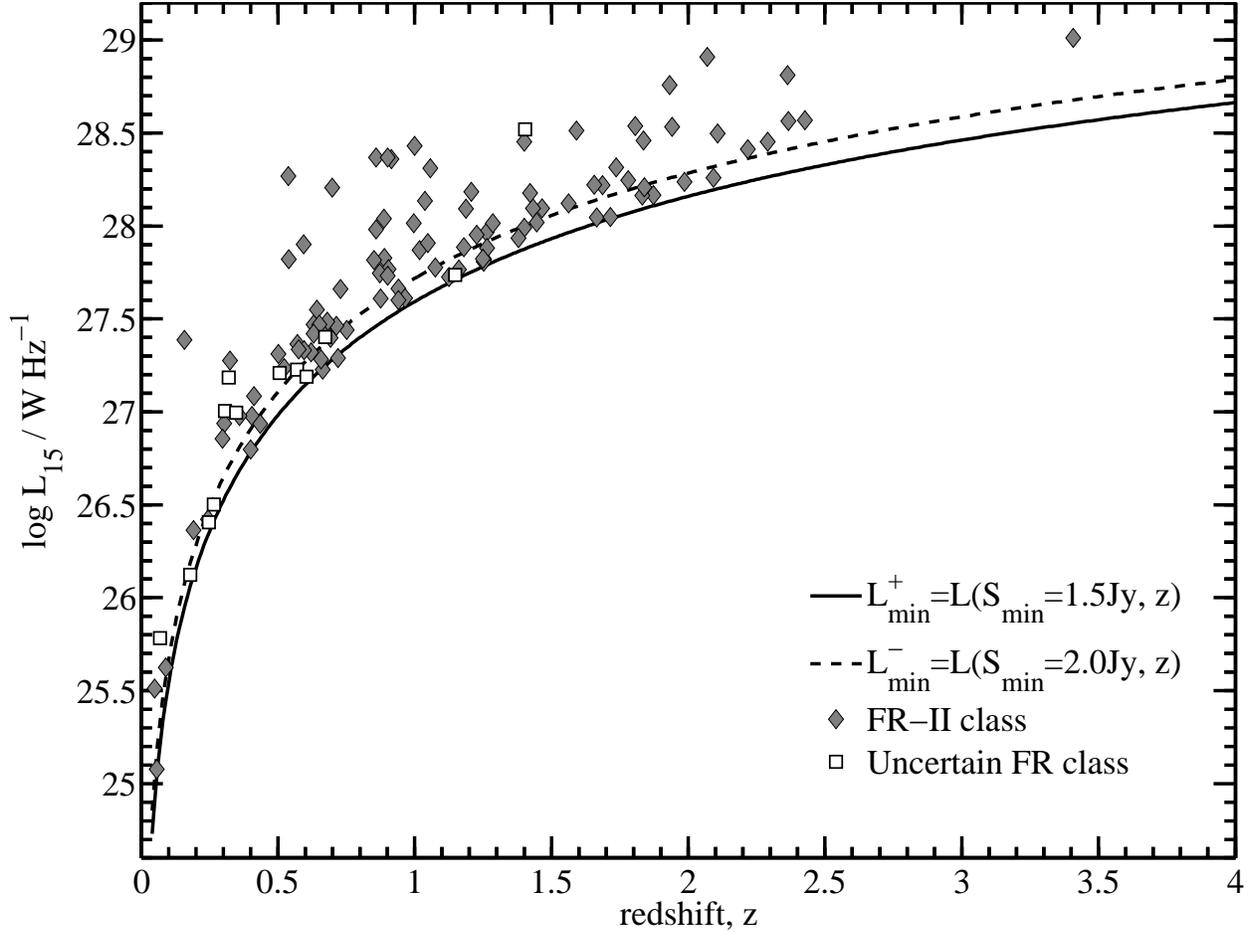}
\end{center}
\caption{\small The luminosity-redshift distribution of
the full MOJAVE sample (omitting 4 known FR~Is and two GPS
sources). Only sources with known redshifts are plotted. We use
diamonds for known FR~II sources and open squares for sources
of uncertain FR class. The solid line corresponds to the
$1.5\Jy$ flux density cutoff for the sources with positive
J2000 declinations and the dashed line corresponds to the
$2\Jy$ flux density cutoff for the sources with negative
declinations, assuming a flat spectral
index.}
\label{fig:Lzplane}
\end{figure}

In Figure \ref{fig:Lzplane} we show the luminosity-redshift
distribution of sample based on the data of \cite{LH05}, as
well as the flux density cutoffs corresponding to the northern
and southern sky regions. The smallest and largest observed
luminosities in our sample are $L_{\min}^{\mathrm{obs}}\approx
1.19\times 10^{25}\WHz$ and $L_{\max}^{\mathrm{obs}}\approx
1.03\times 10^{29}\WHz$ and the redshifts range from
$z_{\min}=0.0491$ to $z_{\max}=3.408$.

\subsection{Missing Redshifts}\label{subsec:missingz}

Despite considerable observational effort, the redshift
information on the MOJAVE sample is incomplete, because of the
featureless optical spectra of several blazars, and
weak/obscured optical counterparts. We address this problem by
building a pool of redshifts from sources which have known
redshifts and flux densities within $0.15\Jy$ of the source
with the unknown redshift. We then randomly select a redshift
from this pool to be used as the redshift for that source.
Alternatively, one could randomly select redshifts from the
entire pool of 102 sources for known FR~II sample (115 for the
full sample), however, we chose the former method because of
the large range of luminosity and redshift spanned by the
sample. Because there is only one missing redshift in the known
FR~II sample, the number of possible redshift combinations is
only 13 for the known FR~II sample, compared to $4.3\times
10^{13}$ for the full sample. In the discussion that follows,
we use 13 (for the known FR~II sample) and 1000 (for the full
sample) realizations of the randomized redshifts to determine
the statistical errors on our best fit model parameters arising
from missing redshift information.

\section{Method}\label{sec:Method}

\subsection{Parameterized Luminosity Function}\label{sec:PLF}

The differential luminosity function of a population of objects
is defined as the number of objects per unit co-moving volume
per unit luminosity interval, i.e.,
\begin{equation}\label{eqn:LFDef}
\phi(\mathcal{L},z)=\frac{d^2N(\mathcal{L},z)}{dV\,d
\mathcal{L}},
\end{equation}
where $N$ is the number of objects of luminosity $\mathcal{L}$
found in the co-moving volume $V$ at redshift $z$. Studies of
flux-limited AGN samples using the $<\mathrm{V/V_{max}}>$ test,
including MOJAVE \citep{Arsh06}, indicate that the RLF
generally evolves with redshift. Without losing generality, we
can write the RLF as
\begin{equation}\label{eqn:LFParts}
\phi(\mathcal{L},z)=\phi_0(\mathcal{L})f_{\mathrm{ev}}
(\mathcal{L},z)
\end{equation}
where $\phi_0(\mathcal{L})$ is the local ($z \simeq 0$) RLF and
$f_{\mathrm{ev}}(\mathcal{L}, z)$ is the evolution function.

For the intrinsic RLF, we adopt a parameterization in which the
local RLF is a simple power law of the form
\begin{equation}\label{eqn:LFPLEnoev}
\phi({\cal L})=
\cases{{\displaystyle \frac{n_0}{L_*}\left(\frac{{\cal L}}
{L_*}\right)^\alpha,} & ${\cal L}_1<{\cal L}<{\cal L}_2$, \cr
0, & elsewhere,}
\end{equation}
where $L_*$ is an arbitrary constant with units of luminosity
and $n_0$ is a normalization constant. In this paper we will
use $L_*=10^{27}\WHz$.

Traditionally, the evolution (in the simplest cases taken to be
luminosity-independent) has been parameterized in two popular
forms: a power-law evolution of the form $(1+z)^k$, or an
exponential evolution of the form $\exp[k\,\tau(z)]$ where
$\tau(z)$ is the look-back time. Other studies
\citep[e.g.,][]{Willott98} have used 1- or 2-tailed Gaussian
redshift dependencies. We were not able to successfully fit the
MOJAVE data using these parameterizations. In particular, in
several cases such parameterizations predicted a large spike in
the number of low-redshift sources, which is not the case for
the MOJAVE sample. Instead, we found that a good fit to the
data could be obtained using the following
luminosity-independent density evolution function:
\begin{equation}\label{eqn:ModelDEDef}
f_{\mathrm{ev}}(\mathcal{L}, z)=f_D(z)\equiv
z^m\exp\left[-\frac{1}{2}\left(\frac{z-z_0}{\sigma}\right)^2
\right],
\end{equation}
where $m$, $z_0$ and $\sigma$ are free parameters of the model.
Note that this function does not reduce to $f_D(z)=1$ at $z=0$;
we therefore assume that the model evolution function is valid
for a range of redshifts $z_1 < z < z_2$. Combining equations
(\ref{eqn:LFParts}) through (\ref{eqn:ModelDEDef}), our
intrinsic model RLF becomes
\begin{equation}\label{eqn:LFLDE}
\phi(\mathcal{L},z) = {\displaystyle
\frac{n_0}{L_*}\left(\frac{\mathcal{L}}{L_*}\right) ^\alpha
z^m\exp\left[-\frac{1}{2}\left(\frac{z-z_0}{\sigma}\right)^2
\right]},
\end{equation}
which is valid over the domain
\begin{equation}\label{eqn:ValidityDomain}
\mathcal{L}_1<\mathcal{L}<\mathcal{L}_2\quad\mbox{and}\quad z_1
< z < z_2.
\end{equation}

Because the luminous jet material is moving with a speed
comparable to $c$ (bulk Lorentz factor $\gamma >> 1$), its
observed monochromatic luminosity will be boosted as
\begin{equation}\label{eqn:LDPLB}
L=\delta^p\mathcal{L},
\end{equation}
where $\mathcal{L}$ is the luminosity in the rest frame,
$p=2-\alpharad$ for continuous jet emission, $\alpharad$ is the
spectral index, and $\delta$ is the kinematic Doppler factor
defined as
\begin{equation}\label{eqn:DPLfac}
\delta=\left(\gamma-\sqrt{\gamma^2-1}\cos{\theta}\right)^{-1},
\end{equation}
where $\gamma=\left(1-\beta^2\right)^{-1/2}$ is the Lorentz
factor and $\beta=v/c$ is the velocity of the emitting plasma.
If the viewing angle to the jet lies within the range
$0\arcdeg\leq\theta\leq90\arcdeg$ and
$\gamma_1\leq\gamma\leq\gamma_2$, then the possible Doppler
factors range from
\begin{equation}\label{eqn:DPLRangeMin}
\delta_{\min}=1/\gamma_2
\end{equation}
to
\begin{equation}\label{eqn:DPLRangeMax}
\delta_{\max}=\gamma_2+\sqrt{\gamma_2^2-1}\;.
\end{equation}

If the intrinsic luminosity $\mathcal{L}$ is Doppler boosted as
in equation (\ref{eqn:LDPLB}), then the distribution of the
observed luminosities $L$ will be different from the
distribution of the intrinsic luminosities. Following the
approach used by \citet{L03}, we derive the form of the
observed RLF of the Doppler beamed sources as
\begin{equation}
\label{eqn:DBLF} \Phi(L,z)={\displaystyle
\frac{n_0}{L_*}\left(\frac{L}{L_*}\right)^\alpha
f_D(z)\int_{\delta_1(L)}^{\delta_2(L)}P_{\delta}(\delta)
\delta^{-p(\alpha+1)}\,d\delta},
\end{equation}
where $P(\delta)$ is the probability density function for
$\delta$. This model function is valid over the domain
\begin{equation}\label{eqn:LumDomain}
L_1<L<L_2\quad\mbox{and}\quad z_1 < z < z_2,
\end{equation}
where
\begin{equation}\label{eqn:LminLmaxDef1}
L_1\equiv\delta_{\min}^p\mathcal{L}_1
\end{equation}
and
\begin{equation}\label{eqn:LminLmaxDef2}
L_2\equiv\delta_{\max}^p\mathcal{L}_2.
\end{equation}

In equation (\ref{eqn:DBLF}), the limits of integration
$\delta_1(L)$ and $\delta_2(L)$ are given by
\begin{equation}\label{eqn:DeltaLimitsD1}
\delta_1(L)=\min\left\{\delta_{\max},\max\left\{\delta_{\min},
\left(L/\mathcal{L}_2\right)^{1/p}\right\}\right\}
\end{equation}
and
\begin{equation}\label{eqn:DeltaLimitsD2}
\delta_2(L)=\max\left\{\delta_{\min},\min\left\{\delta_{\max},
\left(L/\mathcal{L}_1\right)^{1/p}\right\}\right\}.
\end{equation}
with $\delta_{\min}$ and $\delta_{\max}$ given by equations
(\ref{eqn:DPLRangeMin}) and (\ref{eqn:DPLRangeMax}). The
probability density function for $\delta$ is
\begin{equation}\label{eqn:PdeltaGeneral}
P_{\delta}(\delta)=\delta^{-2}\int^{\gamma_2}_{f(\delta)}
\frac{P_{\gamma}(\gamma)}{\sqrt{\gamma^2-1}}\,d\gamma,
\end{equation}
where $P_{\gamma}(\gamma)$ is the probability density function
for $\gamma$ and the lower limit of integration is given in
equation (A6) of \citet[][]{L03}. According to the previous
results of \citet{LM97}, we adopt a power-law form of
$P_{\gamma}(\gamma)$ with index $k$:
\begin{equation}\label{eqn:Pgamma}
P_{\gamma}(\gamma)=C\gamma^k
\end{equation}
for $\gamma_1<\gamma<\gamma_2$, where $C$ is a normalization
constant.

For computational purposes we express $P_{\delta}(\delta)$
using beta functions (see Appendix \ref{app:PDBeta}, eq.
\ref{eqn:A2PDelta}) as
\begin{equation}\label{eqn:PdeltaBeta}
P_\delta \left( \delta \right) =\frac C{2\delta ^2}\left\{ B
\left(1-\frac 1{\gamma _2^2},\frac 12,-\frac k2\right) -B\left(
1-\frac 1{f^2\left( \delta \right) },\frac 12,-\frac k2\right)
\right\}.
\end{equation}

\subsection{Maximum Likelihood Method}\label{sec:MLMethod}

From equation (\ref{eqn:DBLF}) it is apparent that the model
parameters ($\alpha, m, z_0, \sigma$) of the Doppler-beamed RLF
are the same as the parameters of the intrinsic RLF. Therefore,
we can find the parameters of the intrinsic RLF by fitting the
Doppler-beamed RLF to the observed data. For this purpose we
use the maximum likelihood method of \citet{Mar83}, which
attempts to minimize the function
$S=-2\ln(\mathrm{Likelihood})$. The integral in $S$ \citep[eq.
2 of][]{Mar83} should be equal to the sample size $N$ for a
good fit. Therefore, we must minimize
\begin{equation}\label{eqn:LogLikelihood2}
S(\alpha, m, z_0, \sigma)=-2\sum_{i=1}^N
\ln\left[\Phi(L_i,z_i)\right]+2N
\end{equation}
and normalize $\Phi(L,z)$ such that
\begin{eqnarray}
\label{eqn:SampleSize1}
N & = & f_{\Omega}^+\int_{z_1}^{z_2}dz \, \frac{dV}{dz}
\int_{\max\{L_1, L^+_{\min}(z)\}}^{L_2}dL \,  \Phi(L,z)
\nonumber \\
& + & f_{\Omega}^-\int_{z_1}^{z_2}dz \, \frac{dV}{dz}
\int_{\max\{L_1,L^-_{\min}(z)\}}^{L_2}dL \,  \Phi(L,z),
\end{eqnarray}
where $N$ is the sample size, and
$f_{\Omega}^+\approx6.00912/4\pi$ and
$f_{\Omega}^-\approx2.08012/4\pi$ are the fractional area of
the sky available to the survey (in this section the ``$+$''
superscript refers to the northern sky area while the ``$-$''
superscript refers to the southern sky area: $0\arcdeg < \delta
\leq -20\arcdeg$). In equation (\ref{eqn:SampleSize1}) we take
into account that in the MOJAVE sample we have two different
non-overlapping sky areas, each with its own flux density
limit: $S^+_{\min}=1.5\Jy$ and $S^-_{\min}=2.0\Jy$. The
$L^+_{\min}(z)$ and $L^-_{\min}(z)$ in the equation
(\ref{eqn:SampleSize1}) are the monochromatic luminosity limits
corresponding to the flux density limits of the survey:
\begin{equation}\label{eqn:Lminz}
L^\pm_{\min}(z)=4\pi S^\pm_{\min}D_L^2(z)(1+z)^
{-(1+\alpharad)},
\end{equation}
where $D_L(z)$ is the luminosity distance. To minimize
$S(\alpha, m, z_0, \sigma)$ we use the ``amoeba'' algorithm
from \citet{NRC92}.

Other parameters of the model, such as the redshift limits
($z_1$ and $z_2$), luminosity limits ($\mathcal{L}_1$ and
$\mathcal{L}_2$), power law exponent $k$ of the Lorentz factor
distribution (eq. \ref{eqn:Pgamma}) and its range of possible
values $[\gamma_1,\gamma_2]$ are taken as fixed a-priori, and
are not included in the set of optimized parameters. Some of
these parameters (e.g., $\mathcal{L}_1$ and $k$) are poorly
constrained (for reasons explained at the end of this section),
while others can be estimated from the data directly, as
follows.

The sources in both the FR~II only sample as well as in the
full sample span a broad range of redshifts from
$z_{\min}=0.0491$ to $z_{\max}=3.408$. Later in this section we
show the upper limit of the intrinsic luminosity of the parent
population to be about $\mathcal{L}_2=10^{29}\WHz$. The flux
limit of the MOJAVE survey would allow the detection of
luminous sources with $\mathcal{L} \geq 10^{28}\WHz$ at $z=4$
if their Doppler factors are $\delta \geq 2.15$. The lack of
such sources in the MOJAVE sample at redshifts $z \gtrsim 3.4$
may be of importance in modeling the RLF of the parent
population. Therefore, by extending the range of redshifts to
$z=4$, we allow more freedom in the optimization procedure. In
addition, this will ensure that we do not exclude the
statistical possibility of some objects at higher redshift.
Because of this we slightly extend the redshift range and set
\begin{equation}\label{eqn:z1z2range}
z_1=0.04\quad\mbox{and}\quad z_2=4.
\end{equation}
Extending the upper redshift limit to higher values should not
have a large effect on the model RLF, since the flux cutoff of
the survey will make the available comoving volume very small
at high $z$ \citep[see, e.g.,][]{Willott98}. In addition,
because of the above mentioned lack of sources above $z \sim
3.4$, the optimization process will constrain the evolution
function (eq. \ref{eqn:ModelDEDef}) to vanish rapidly at larger
redshifts.

Previous studies \citep[see, e.g.,][]{Homan06,Cohen07} have
shown that the apparent speeds of powerful AGN jets are closely
related to their bulk flow velocities and VLBI core properties.
Since the MOJAVE sample contains powerful AGN with highly
core-dominated radio structures \citep{Cooper07} and
superluminal jets \citep{Kellermann04}, one might expect the
parent population to have $\gamma_1 \gg 1$. However, the parent
population likely contains sources with much lower jet speeds,
and indeed, \citet{Cohen07} estimates that the jet speed in
Cygnus~A (one of the sources in our subsamples) is
$0.59<\beta<0.68$. Other authors \citep[e.g.,][]{Wardle97}
obtain similar estimates  for jet speeds in kiloparsec scales
outflows ($\beta \geq 0.6$). Assuming no strong deceleration of
the jets in FR~II sources, we adopt the value of
$\beta_{\min}=0.6$, or $\gamma_1=1.25$, for the minimum jet
speeds in the parent population. We will discuss the effects of
this choice on our model LF in subsection
\ref{subsec:ModelPar}. Using recent observational data,
\citet{Cohen07} find that for the MOJAVE sample,
$\gamma_{\max}\approx32$. \citet{LM97} find, using Monte Carlo
simulations, that a power-law exponent of the Lorentz factor
distribution in the range $-1.5 \lesssim k \lesssim -1.75$
provide a reasonable fit to the CJ-F survey \citep{Taylor96}, a
comparable radio-loud blazar sample. In this paper we consider
the following range of possible Lorentz factors and the
exponent $k$:
\begin{equation}\label{eqn:gammapar}
\gamma_1=1.25,\quad\gamma_2=32\quad\mbox{and}\quad k=-1.5.
\end{equation}

We can estimate the lower and upper limits for the intrinsic
luminosity as follows. First, from equations
(\ref{eqn:DPLRangeMin}), (\ref{eqn:DPLRangeMax}) and
(\ref{eqn:gammapar}) we obtain
\begin{eqnarray*}
\delta_{\min}=1/\gamma_2= 0.031\quad\mbox{and}\quad
\delta_{\max}=\gamma_2+\sqrt{\gamma_2^2-1}\approx 64 \nonumber
\end{eqnarray*}
and we can apply the equation (\ref{eqn:LDPLB}) to the observed
luminosity range in our sample: $L_{\min}^{\mathrm{obs}}\approx
1.19\times 10^{25}\WHz$ and $L_{\max}^{\mathrm{obs}}\approx
1.03\times 10^{29}\WHz$. For example, in the extreme case where
the range of intrinsic luminosities ($\mathcal{L}$) is
maximized, we have $L_{\min}^{\mathrm{obs}}=\mathcal{L}_1
\delta_{\max}^2$ and $L_{\max}^{\mathrm{obs}}=\mathcal{L}_2
\delta_{\min}^2$ and we obtain $\mathcal{L}_1\approx 3.31\times
10^{21}\WHz$ and $\mathcal{L}_2\approx 1.05\times 10^{32}\WHz$.
In reality, there is a very low probability of having such
extreme values in the MOJAVE sample (see, e.g.,
\citealt*{Cohen07}). To fine tune this range, we initially fit
the data using the values given above. We used the parameters
of the resulting fitted RLF to produce a large population of
sources via Monte Carlo simulations. We examined the intrinsic
luminosity distribution of a simulated flux-limited sample to
see if many sources had intrinsic luminosities near the value
of $\mathcal{L}_1$. If all sources were well above this value,
we adjusted $\mathcal{L}_1$ upward incrementally until we
obtained a tight fit of the simulated distribution of the
intrinsic luminosities to the initial range used in that
particular step.  A similar procedure was applied for the upper
limit $\mathcal{L}_2$. In this manner we found that for the
known FR~II sample
\begin{equation}\label{eqn:InrinsicLRange1}
\mathcal{L}_1=10^{22.2}\approx 1.58\times10^{22}\WHz
\end{equation}
and
\begin{equation}\label{eqn:InrinsicLRange2}
\mathcal{L}_2=10^{29.1}\approx 1.26\times10^{29}\WHz
\end{equation}
provided a good fit to the simulated intrinsic luminosity
histogram. Similarly, for the full sample we find the following
limits for the intrinsic luminosities
\begin{equation}\label{eqn:InrinsicLRange1b}
\mathcal{L}_1=10^{21.6}\approx 3.98\times10^{21}\WHz
\end{equation}
and
\begin{equation}\label{eqn:InrinsicLRange2b}
\mathcal{L}_2=10^{29.2}\approx 1.58\times10^{29}\WHz .
\end{equation}
For these Monte Carlo simulations we have used the 64-bit
random number generator (RNG) of \citet{Marsaglia04} as many
commonly used algorithms \citep[e.g., Numerical
Recipes][]{NRC92} lack the necessary resolution for generating
deviates that a needed to span the wide range of luminosities
found in equations
(\ref{eqn:InrinsicLRange1})--(\ref{eqn:InrinsicLRange2b}).

Substituting the above range of intrinsic luminosities into
equations (\ref{eqn:LminLmaxDef1}) and
(\ref{eqn:LminLmaxDef2}), we obtain a theoretical range for the
observed luminosities: $L_1\approx 1.55\times 10^{19}\WHz$ and
$L_2\approx 5.15\times 10^{32}\WHz$ for the known FR~II sample
and $L_1\approx 3.89\times 10^{18}\WHz$ and $L_2\approx
6.49\times 10^{32}\WHz$ for the full sample. These ranges are
much larger than the observed range of luminosities in the
MOJAVE sample (see $L_{\min}^{\mathrm{obs}}$ and
$L_{\max}^{\mathrm{obs}}$ above). Because the model RLF is not
well determined outside the observed luminosity range, we adopt
a conservative approach and adopt the following validity range
for the luminosities of the observed (beamed) RLF:
$L_1=10^{25}\WHz$ and $L_2=1.1\times10^{29}\WHz$.

Using the values of $\mathcal{L}_1$ and $\mathcal{L}_2$ from
equations (\ref{eqn:InrinsicLRange1}) and
(\ref{eqn:InrinsicLRange2}), we find that the MOJAVE cutoff
luminosities are too high for us to observe some important
features of the RLF. For example, from Fig. 3 of \cite{L03} it
is evident that we would need to observe below the luminosity
$L_4\approx 6.49\times 10^{25}\WHz$ for the known FR~II sample
($L_4\approx 1.63\times 10^{25}\WHz$ for full sample)
\citep[see][eq. 9]{L03} to probe the region of the RLF that is
most susceptible to the changes in values of the lower
luminosity $\mathcal{L}_1$ of the parent population and
power-law index $k$.  But in our sample we have too few sources
with $L<L_4$. For these reasons we chose to estimate some
parameters of the model from the data as described above, and
not to include them in the set of optimized parameters.

\section{Results}\label{sec:RLFResults}

\subsection{Model Parameters}\label{subsec:ModelPar}

Using our adopted form of density evolution (eq.
\ref{eqn:ModelDEDef}) and parameters from equations
(\ref{eqn:z1z2range}), (\ref{eqn:gammapar}),
(\ref{eqn:InrinsicLRange1}) and (\ref{eqn:InrinsicLRange2}), we
minimized the quantity $S(\alpha, m, z_0, \sigma)$ for 1000
(182 for the known FR~II sample) randomizations of missing
redshifts as described in \S~\ref{subsec:missingz}. For each
fitted parameter, we took the median of the distribution as the
best fit value of the respective parameter. The best fit values
of the model RLF thus obtained are presented in
Table~\ref{tab:BestFitParams}. The error estimates for model
parameters have been obtained using the $\Delta S = 1$ method
\citep[see,][]{Lampton76}. (For a given parameter we maximized
the likelihood function while keeping all other parameters
constant. We then varied this parameter until a variation
$\Delta S = 1$ was obtained.) We have also obtained an
estimation of the error in the parameter due to missing
redshift information at the level of $1\sigma$ from the values
of the parameters for which the fractional cumulative
distribution function was equal to either $0.683$ or $0.317$.
We found that these errors are negligible compared to the
errors computed using the $\Delta S=1$ method. We calculated
the normalization constant $n_0$, space density $\rho$ for
$L>1.3\times 10^{25}$, and parent population $K$ using the best
fit values for the model parameters $\alpha, m, z_0, \sigma$ so
that equation (\ref{eqn:LogLikelihood2}) yielded the sample
size $N=103$ for the known FR~II sample and $N=127$ for the
full sample.  The errors on $n_0$ and $K$ were also calculated
using their cumulative distribution functions as described
above. We have evaluated the goodness-of-fit of our model RLFs
using the two-dimensional Kolmogorov-Smirnov (K-S) test as
described in \citet{NRC92}. The Kolmogorov-Smirnov probability
($\mbox{P}_{\mathrm{KS}}$) is a \emph{p-value} that shows the
probability of observing a K-S test statistic
($\mbox{D}_{\mathrm{KS}}$) as large or larger than observed one
and can be used to reject a model if its value is too small. We
will accept a model if $\mbox{P}_{\mathrm{KS}}\geq 0.2$. The
K-S probabilities for our models indicated good fits to the
data with Monte Carlo realizations of missing redshifts,
ranging from 0.67 to 0.79 for the known FR~II sample (with
$\mbox{D}_{\mathrm{KS}}$ from 0.067 to 0.079) and  from 0.40 to
0.97 for the full sample (with $\mbox{D}_{\mathrm{KS}}$ from
0.048 to 0.089).

From Table~\ref{tab:BestFitParams} we can see that while the
slope for the intrinsic luminosity distribution of the FR~II
only sample is slightly shallower than the slope of the full
sample, the parameters of the density evolution functions of
the two samples agree to within $1\sigma$, suggesting that the
objects of uncertain FR class (9 BL~Lacs, 9 BL~Lac/HPQ sources
and 6 sources without optical classification) may actually be
of the FR~II class.

The median of average space densities (for $L>1.3\times
10^{25}\WHz$) computed for 13 (1000 for the full sample)
randomizations of the unknown redshifts is $\approx
1580\GpcCubed$ for the known FR~II sample and $\approx
4390\GpcCubed$ for the full sample. For the full sample the
space density is quite large. We speculate that this may
indicate that the RLF has a different slope for lower
luminosities, but the lack of low luminosity sources in our
sample does not allow us to verify this hypothesis. We can also
explain this increase in the space density for the full sample
by an underestimation of the lower intrinsic luminosity
$\mathcal{L}_1$ (see equation (\ref{eqn:InrinsicLRange1b})) due
to the large number of sources with missing redshifts in the
full sample.

As discussed in subsection \ref{sec:MLMethod}, the value of the
lower limit of the Lorentz factors $\gamma_1$ is not very well
constrained, with some authors \citep[e.g.,][]{Arsh06}
suggesting higher values (e.g., $\gamma_1=3$) than the one
adopted here ($\gamma_1=1.25$). Therefore, to investigate
whether or not the choice of a particular value of $\gamma_1$
has a significant influence on the model RLF, we have repeated
the computations for the FR~II-only sample using $\gamma_1=3$.
While we have obtained a slightly different intrinsic
luminosity range ($\mathcal{L}_1=10^{21.8}\WHz$ and
$\mathcal{L}_2=10^{29}\WHz$), the model RLF parameters are
essentially unchanged: $\alpha=-2.55 \pm 0.06$, $m=1.48 \pm
0.14$, $z_0=1.27 \pm 0.09$, and $\sigma=0.76 \pm 0.10$. This
relative independence of the results on a particular choice of
the lower limit $\gamma_1$ for the Lorentz factors is due to
the insensitivity of the bright end ($L>L_4$) of the beamed RLF
to the values of $\gamma_1$ and $\gamma_2$ \citep[see][Figure
5]{L03}. However, the fact that the luminosity functions in
Figure 5 of \citet{L03} differ strongly at lower luminosities
than $L_4$ means that the choice of $\gamma_1$ potentially
could modify the predicted parent population sizes. Indeed, for
the FR~II only sample with $\gamma_1=3$, we obtain
$\mbox{K}=(1.7 \pm 0.1)\times 10^{10}$, which is about three
times larger than the parent population predicted by the RLF
computed with $\gamma_1=1.25$, but this could be due to the
lower value of $\mathcal{L}_1$ obtained when $\gamma_1=3$.

\begin{figure}[htbp]
\begin{center}
\includegraphics[angle=-90,scale=0.65]{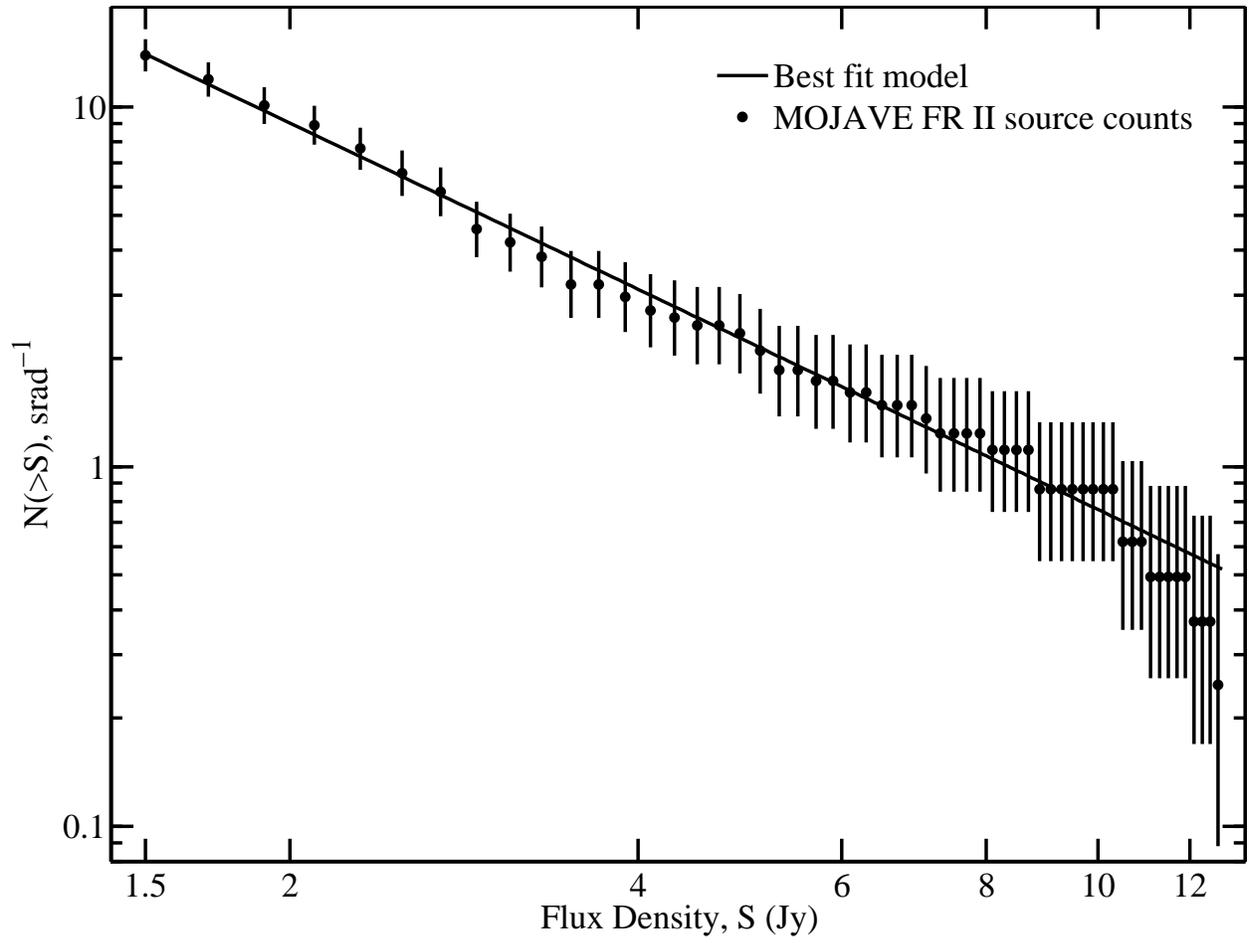}
\end{center}
\caption{\small Integral source count of the known FR~II
sample $N(>S)$ per unit of solid angle.}
\label{fig:SrcCounts}
\end{figure}

In Figure \ref{fig:SrcCounts} we present the integral source
counts $N(>S)$ per unit of solid angle for the observed data
(known FR~II sample), and as predicted by our fitted RLF after
it is beamed. The $1\sigma$ error bars in this and subsequent
figures are computed according to Poisson statistics using the
method of \citet{Gehrels86}.

\subsection{Redshift Distribution}

\begin{figure}[htbp]
\begin{center}
\includegraphics[angle=-90,scale=0.65]{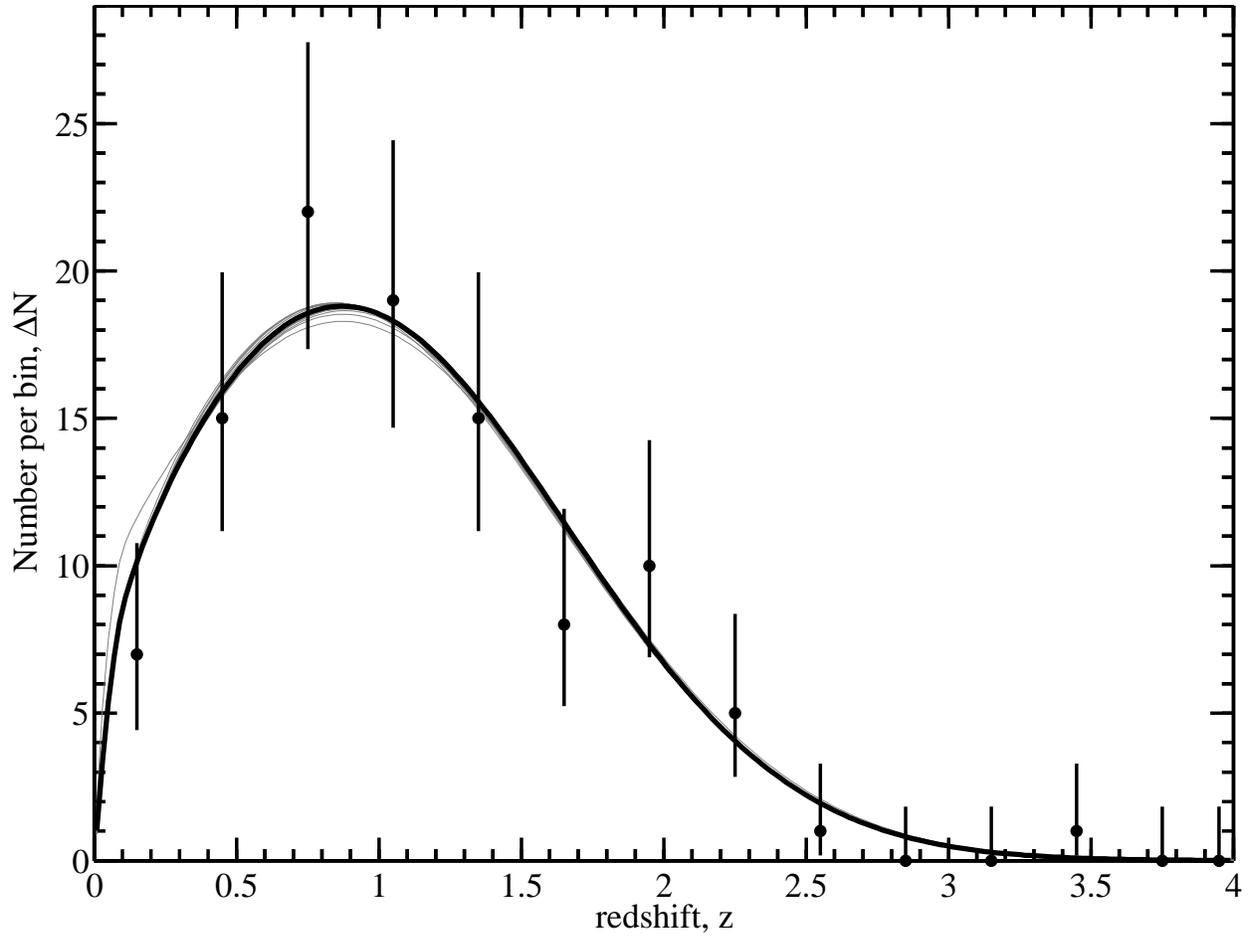}
\end{center}
\caption{\small Plot of model redshift distributions for
each redshift randomization (thin light gray lines), best fit
model distribution (thick black line) and the observed redshift
distribution of the MOJAVE known FR~II sample (filled circles
with error bars corresponding to $1\sigma$ confidence level).}
\label{fig:randzdistrib}
\end{figure}

In Figure \ref{fig:randzdistrib} we plot the binned redshift
distribution and the associated $1\sigma$ error bars for the
known FR~II sample (with the missing redshifts replaced with
the averages of the ``redshift pools'' as described in
\S~\ref{subsec:missingz}). The solid line represents the
predicted redshift distribution for our best fit model, while
the faint gray lines show the distributions for the 13
randomizations of the missing redshifts. We can see that while
the missing redshift information creates a tangible uncertainty
in the redshift distribution, we obtain a reasonably good
overall fit to the data.

\subsection{Radio Luminosity Function}\label{sec:RLFresults}

We use the method of \citet{PC00} to construct the observed
luminosity function. In this method, we compute the value of
the RLF in a bin with a luminosity interval $L_{\min}$ and
$L_{\max}$ and a redshift interval $z_{\min}$ and $z_{\max}$
as:
\begin{equation}
\label{eqn:BinLogLF}
\Phi_{\mbox{est}}=\frac{N}{\int_{z_{\min}}^{z_{\max}}
\int_{L_{\min}(z)}^{L_{\max}}\frac{dV}{dz}dzdL}
\end{equation}
and its uncertainty:
\begin{equation}\label{eqn:DBinLogLF}
\delta\Phi_{\mbox{est}}=\frac{{\delta}N}{\int_{z_{\min}}^
{z_{\max}}\int_{L_{\min}(z)}^{L_{\max}}\frac{dV}{dz}dzdL} ,
\end{equation}
where $N$ is the number of objects in the bin and $\delta{N}$
its uncertainty. $L_{\min}(z)$ is the minimum luminosity within
the bin at which we can still detect an object. In equations
(\ref{eqn:BinLogLF}) and (\ref{eqn:DBinLogLF}) we have switched
the order of integration compared to the original formulation
of \citet{PC00}.

We use these same equations to compute the binned (i.e.,
averaged over a luminosity-redshift bin) model RLF.  We believe
this is the most robust way to compare the observed and model
RLF when binning is involved. We use equations
(\ref{eqn:SampleSize1}) to compute the effective ``number of
sources'' $N$ in the luminosity-redshift bin of interest (i.e.,
we replace $z_1$, $z_2$, $L_1$ and $L_2$ in eq.
\ref{eqn:SampleSize1} with $z_{\min}$, $z_{\min}$, $L_{\min}$
and $L_{\max}$ of the luminosity-redshift bin of interest).

\begin{figure}[htbp]
\begin{center}
\includegraphics[angle=-90, scale=0.65]{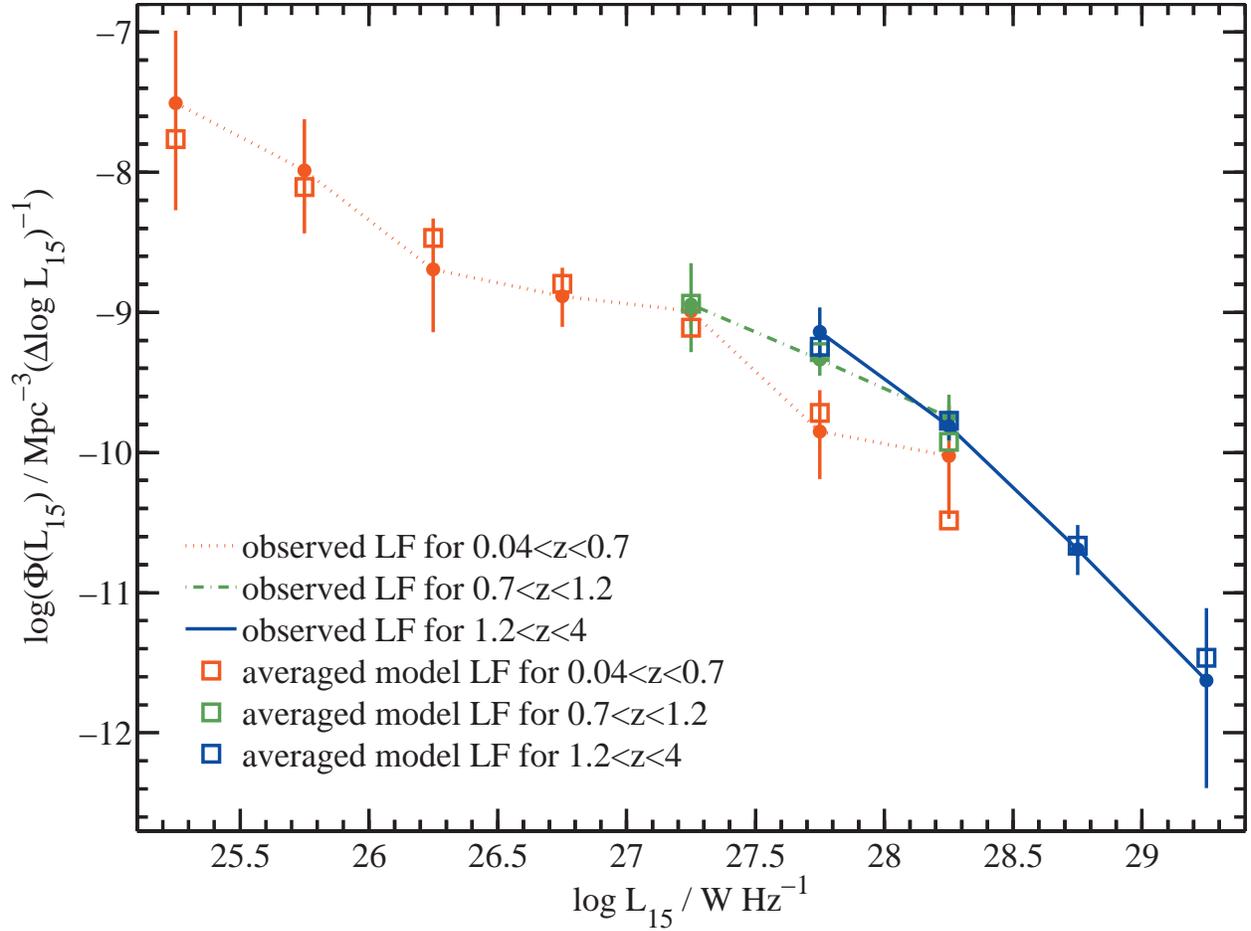}
\end{center}
\caption{\small The observed binned RLF of the MOJAVE
known FR~II sample (filled circles and lines) and the binned
model RLF (open squares) for three redshift intervals:
$0.04<z<0.7$ (red, dotted line), $0.7<z<1.2$ (green,
dot-dashed line), and $1.2<z<4$ (blue, solid line). The
luminosity bin width is $\Delta \log L=0.5$.}
\label{fig:BinLF}
\end{figure}

\begin{figure}[htbp]
\begin{center}
\includegraphics[angle=-90,scale=0.65]{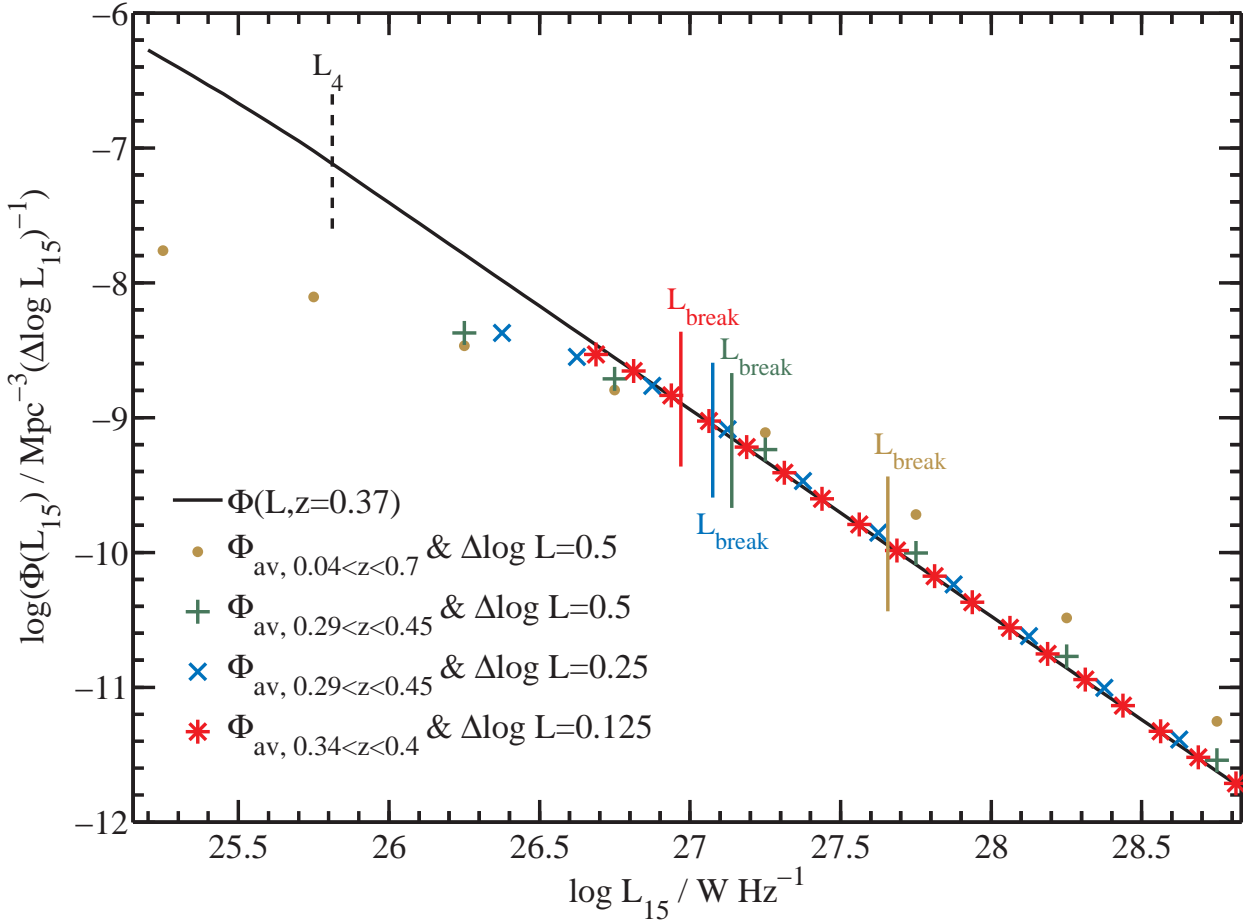}
\end{center}
\caption{\small Differential model LF $\Phi(L,z)$ and
differential LFs averaged over bins of varying size. The best
agreement between the differential LF and its binned version is
obtained for smaller bin sizes. For large bin sizes the binned
(or averaged over bins) LF flattens for lower luminosities. The
vertical lines show the ``break'' luminosities for each LF:
$\log L_{\mathrm{break}}=\log L_{\min}^-(S_{\min }^-=2\Jy,z_{\max}^
{\mathrm{bin}})+(\Delta \log L)/2$.}
\label{fig:LFbreak}
\end{figure}

The fitted and observed RLFs for the MOJAVE sample are
presented in the Figure \ref{fig:BinLF}. It is apparent that
the averaged model RLF provides a good fit to the sample data.
At first glance, it would also appear that both the fitted and
observed RLFs obey a broken power-law that steepens at higher
luminosities. While a beamed LF can have different slopes for
different luminosity intervals \citep[see e.g.,][]{Urry84,L03},
this is does not explain the observed break in the observed
(beamed) LF (e.g., because $L_4$ is too low for our sample; see
Figure \ref{fig:LFbreak}). Based on this \citealt*{Arsh06}
claimed that a double power-law \emph{intrinsic} LF is needed
to describe the observed (beamed) LF. However, we find that
this is an artifact of the binning method.

\begin{figure}[htbp]
\begin{center}
\includegraphics[angle=-90,scale=0.65]{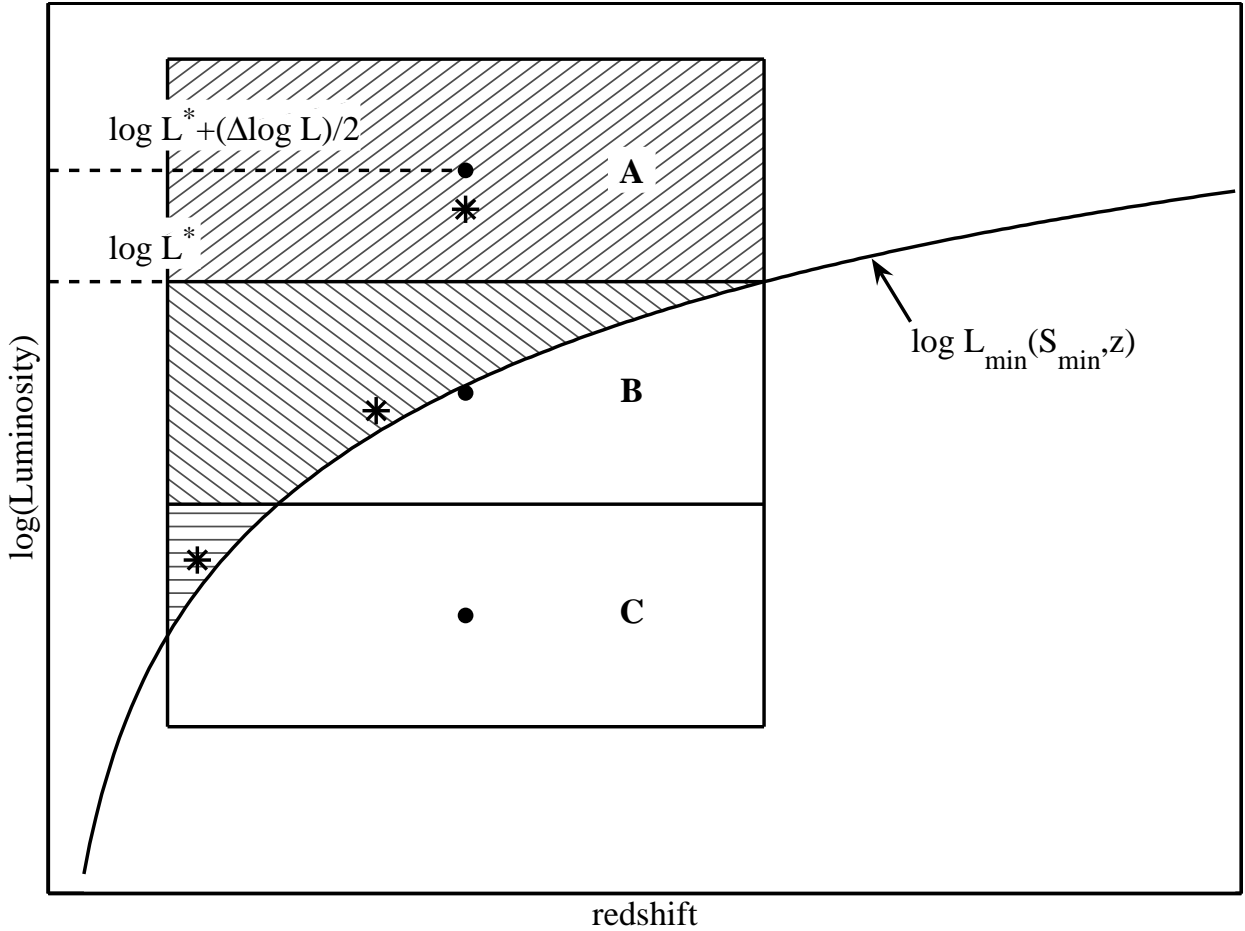}
\end{center}
\caption{\small Illustration of distortions induced by
binning and flux cutoff on the binned LF. The curved line
indicates the flux cutoff: $L_{\min}(S_{\min},z)$. Rectangles
represent bins in the $\log (\mbox{Luminosity})$---redshift
plane. The value of the binned LF is computed as the average of
the differential LF over hatched regions of the bins. Filled
circles represent the position of the center of the bins (the
luminosity coordinates of these centers were used to plot the
binned LF in Figure \ref{fig:BinLF} and in Figure
\ref{fig:LFbreak}). The asterisks show the position of the
``center of mass'' of the hatched regions for the bins assuming
a $\Phi(L,z)=L^{-2.5}$. Because of the power-law distribution
of luminosities ($L^{-2.5}$), the center of mass for the bin
``A'' lies below its geometrical center. If, in addition,
evolution is present, then the center of mass will shift along
the redshift axis as well.}
\label{fig:LFbreak2}
\end{figure}

\begin{figure}[htbp]
\begin{center}
\includegraphics[angle=-90,scale=0.65]{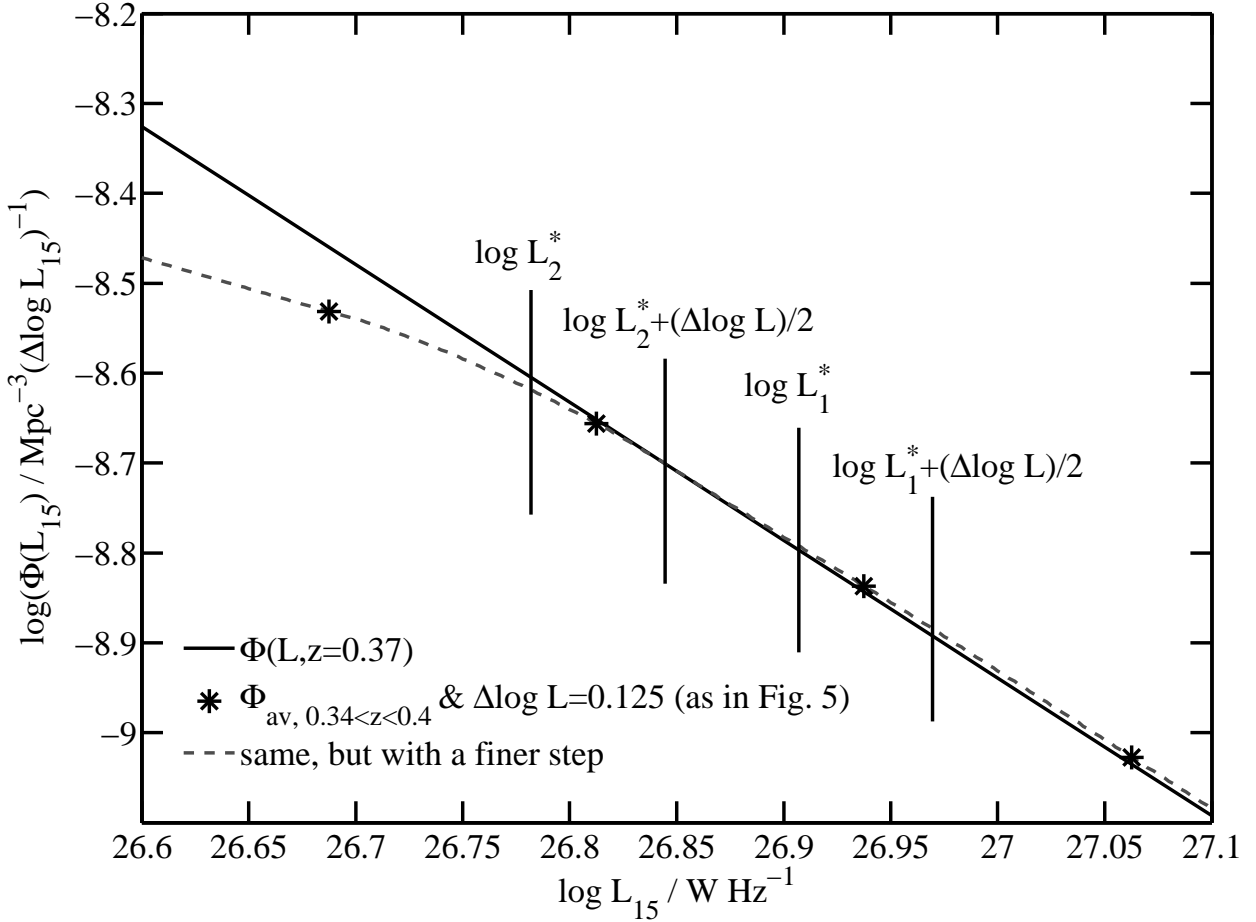}
\end{center}
\caption{\small Differential model LF $\Phi(L,z)$
(continuous line) and the differential LF averaged over bins of
size $\Delta z = 0.06\quad(0.34<z<0.4)$  and
$\Delta\log L=0.125$ (the binned LF represented by red
asterisks in Figure \ref{fig:LFbreak}) (dashed line and
asterisks). Asterisks are placed at the same positions as in
Figure \ref{fig:LFbreak}. The vertical lines show the ``break''
luminosities $\log L_1^{\ast}+(\Delta\log L)/2$ and
$\log L_2^{\ast}+( \Delta\log L)/2$ where
$L_1^{\ast}=L_{\min}^+(S_{\min }^+=1.5\Jy, z=0.4)$ and
$L_2^{\ast}=L_{\min}^-(S_{\min}^-=2\Jy,z=0.4)$.}
\label{fig:LFbreak3}
\end{figure}

In Figure \ref{fig:LFbreak} we plot the Doppler beamed
differential model RLF for the known FR~II sample as well as
its average over bins of varying sizes. We can see that our
differential Doppler beamed RLF is in fact very close to a
simple power law RLF, with only a slight flattening for
$L<L_4$. Because of the large bins, the agreement between the
binned RLF (Figure \ref{fig:BinLF}) and the differential RLF
(Figure \ref{fig:LFbreak}) is apparent only at high
luminosities. This can be improved by using smaller bin sizes,
but at the expense of larger Poisson errors. For a steep RLF
with strong evolution across a bin, large bins can create
apparent breaks in the observed RLFs when these bins intersect
the luminosity cutoff of the sample (see eq. \ref{eqn:Lminz}).
This is because the averages of RLFs computed over parts of the
bins above the luminosity cutoff will be very different from
the averages computed over whole bins (see Figure
\ref{fig:LFbreak2}). A superior way to plot a binned LF would
be to use ``centers of mass'' of the bins over which averaging
is done instead of their geometrical centers but unfortunately,
this is almost impossible to accomplish without a priori
knowledge of the luminosity function slope and evolution
parameters. We can see from Figure \ref{fig:LFbreak2} that
``the center of mass'' of the bins is different from the
geometrical center, even for bins lying entirely above the flux
cutoff (i.e., bin ``A'' in Figure \ref{fig:LFbreak2}). The
presence of a redshift dependence in the luminosity function
will shift the ``center of mass'' of the bins along the
redshift axis as well. The ``chopped'' bins will not be
centered around the same redshifts as the whole bins, and
therefore, the average of the RLF computed over a ``chopped''
bin should actually belong to a RLF computed at a different
cosmological epoch. That is, by assuming that the value of the
RLF at the center of the bin is equal to the average RLF
computed over a smaller part of the bin (e.g., the hatched area
of the bin ``C'' in Figure \ref{fig:LFbreak2}, which may even
not contain the center of the bin) one can introduce a large
error in the case of strongly evolving functions. In Figure
\ref{fig:LFbreak3} we present a zoomed in version of the Figure
\ref{fig:LFbreak} on which, additionally, we plot the ``break''
luminosities.  We can see that the shifting of the ``center of
mass'' towards lower luminosities for ``full'' bins produce a
shift of the LF to the right for $\log L>\log
L_2^{\ast}+(\Delta\log L)/2$. For $\log L<\log
L_2^{\ast}+(\Delta\log L)/2$, the LF flattens because of the
shifting of the ``center of mass'' for the ``chopped'' bins
towards larger luminosities as well as towards smaller
redshifts (see Figure \ref{fig:LFbreak2}). We conclude that the
double power-law that we see in the observed RLF is, therefore,
an artifact created by the effect of flux density cutoff on
steep power law of the intrinsic RLF combined with a strong
evolution of the luminosity function across a bin. This double
power law is not a property of the intrinsic RLF of the MOJAVE
sample as \citet{Arsh06} concluded. In fact, we are able to
obtain good fit using a simple power-law intrinsic RLF.

As previously mentioned in \S~\ref{sec:MLMethod}, the RLF
depends most strongly on the lower luminosity cutoff
$\mathcal{L}_1$ and Lorentz factor distribution power-law index
$k$ at luminosities smaller than $L_4$. At larger luminosities
it appears more like a featureless simple power-law, as
illustrated in Figure \ref{fig:LFbreak}. Indeed, from
mathematical considerations, the slope of the beamed RLF is
expected to be nearly identical to the slope of the unbeamed
(intrinsic) RLF for luminosities between $L_4\approx 6.49\times
10^{25}\WHz$ and $L_8\approx 5\times 10^{29}\WHz$ for the known
FR~II sample and $L_4\approx 1.63\times 10^{25}\WHz$ and
$L_8\approx 4.8\times 10^{29}\WHz$ \citep[see][]{L03,Urry91a}.

\section{Conclusions}\label{sec:Conclusions}

We have analyzed the redshift and flux density distributions of
a complete sample of AGN selected on the basis of
relativistically beamed 15 GHz radio flux density (MOJAVE) to
derive the parent luminosity function of bright radio-loud
blazars. We carried out our analysis on two samples, one
consisting of only the 103 known FR~II class radio sources in
MOJAVE (``the known FR~II sample'') and a ``full sample'' that
added 24 sources of uncertain FR class.

\begin{enumerate}
    \item We find that the observed MOJAVE RLF can be
        well-fit using a Doppler-boosted, single power-law
        intrinsic RLF with slope $\alpha=-2.53 \pm 0.06$
        for the known FR~II sample ($\alpha=-2.65 \pm 0.06$
        for the full sample), and a density evolution
        function of the form $z^m\exp\left[-\frac{1}{2}
        \left(\frac{z-z_0}{\sigma}\right)^2 \right]$, with
        parameters $m=1.4 \pm 0.1$, $z_0=1.29 \pm 0.09$,
        and $\sigma=0.76 \pm 0.09$ for the known FR~II
        sample ($m=1.6 \pm 0.1$, $z_0=1.18 \pm 0.09$, and
        $\sigma=0.80 \pm 0.1$ for the full sample). We
        assumed a power-law Lorentz factor distribution
        with the exponent $k=-1.5$ and $1.25<\gamma<32$.
        Our model is valid over the range $0.04 < z < 4$ in
        redshift, $10^{22.2}\WHz<\mathcal{L}<10^{29.1}\WHz$
        in intrinsic luminosity for the known FR~II sample
        ($10^{21.6}\WHz<\mathcal{L}<10^{29.2}\WHz$ for the
        full sample), and $10^{25}\WHz<L<1.1\times
        10^{29}\WHz$ in observed luminosity.

    \item We find a good agreement between the fitted RLF
        parameters of the two samples, suggesting that the
        objects of uncertain FR class in the MOJAVE sample
        (9 BL~Lacs, 9 intermediate quasar/BL~Lac sources
        and 6 sources without optical classification) may
        in fact belong to the FR~II class.

    \item We have shown that the double power-law shape of
        the \emph{observed} (i.e., beamed) MOJAVE RLF is an
        artifact due to large changes of the evolving RLF
        across a bin and its interaction with the lower
        luminosity cutoff of the survey. We find no
        evidence for a break in the intrinsic blazar RLF
        above $10^{25}\WHz$ at $15\GHz$.
\end{enumerate}

\acknowledgements

The authors wish to acknowledge the contributions of the other
members of the MOJAVE project team: Hugh and Margo Aller,
Tigran Arshakian, Marshall Cohen, Dan Homan, Matthias Kadler,
Ken Kellermann, Yuri Kovalev, Andrei Lobanov, Eduardo Ros, Rene
Vermeulen, and Tony Zensus. We also thank the anonymous referee
for helpful comments and suggestions that improved this paper.

M. Cara wishes to thank Andrzej Lewicki and Michael Sloothaak
for providing additional computational resources at Purdue
University.

This research was supported by NSF grant 0406923-AST, a grant
from the Purdue Research Foundation, and made use of the
following resources:

The NASA/IPAC Extragalactic Database (NED), which is operated
by the Jet Propulsion Laboratory, California Institute of
Technology, under contract with the National Aeronautics and
Space Administration.

The University of Michigan Radio Astronomy Observatory, which
is supported by the National Science Foundation and by funds
from the University of Michigan.

\appendix

\section{Representation of $P_\delta \left( \delta \right) $
using incomplete beta-functions} \label{app:PDBeta}

In order to avoid direct numerical integration, it is
convenient to express the $P_\delta \left( \delta \right) $
through the beta-functions and then use the continued-fraction
representation \citep[see][]{NRC92} for fast computation of the
integral.

If
\[
P_\gamma \left( \gamma \right) =C\gamma ^k
\]
then
\begin{equation}
P_\delta \left( \delta \right) =\delta
^{-2}\int\nolimits_{f\left( \delta \right) }^{\gamma
_2}\frac{P_\gamma \left( \gamma \right) } {\sqrt{\gamma
^2-1}}d\gamma =C\delta ^{-2}\int\nolimits_{f\left( \delta
\right) }^{\gamma _2} \frac{\gamma ^k}{\sqrt{\gamma
^2-1}}d\gamma \equiv C\delta ^{-2}G\left( k,f\left( \delta
\right) ,\gamma _2\right)
\end{equation}
where we have defined
\begin{equation}
G\left( k,z_1,z_2\right) \equiv \int\nolimits_{z_1}^{z_2}
\frac{\gamma ^k}{\sqrt{\gamma ^2-1}}d\gamma .
\end{equation}
Making the substitution $t\equiv \gamma ^2$, this integral can
be rewritten as:
\begin{eqnarray}
G\left( k,z_1,z_2\right)  &=&\frac
12\int\nolimits_{z_1^2}^{z_2^2}
\gamma ^{\frac{k-1}2}\left( t-1\right) ^{-\frac 12}d\gamma \\
&=&\frac 12\int\nolimits_{z_1^2}^{z_2^2}t^{\frac{k+1}2-1}\left(
t-1\right)
^{\frac 12-1}dt  \nonumber \\
&=&\frac i2\left\{ B\left( z_2^2,\frac{k+1}2,\frac 12\right)
-B\left( z_1^2,\frac{k+1}2,\frac 12\right) \right\}
\nonumber
\end{eqnarray}
where
\[
B\left( z,a,b\right) \equiv \int\nolimits_0^zt^{a-1}\left(
t-1\right) ^{b-1}dt
\]
is the incomplete beta function defined for $0\leq z\leq 1$,
$a>0$ and $b>0$. In our problem $z_2>z_1\geq 1$ and therefore
we represent the beta function through the hyper-geometric
function
\begin{equation}
B\left( z,a,b\right) =a^{-1}z^a\,_2F_1\left( a,1-b,a+1;z\right)
\end{equation}
and continue it analytically into the region $\left| z\right|
>1$ using the formula \citep[see][eq. (e.6)]{Landau89}:
\begin{eqnarray}
_2F_1\left( \alpha ,\beta ,\gamma ;z\right)  &=&\frac{\Gamma
\left( \gamma \right) \Gamma \left( \beta -\alpha \right)
}{\Gamma \left( \beta \right) \Gamma \left( \gamma -\alpha
\right) }\left( -z\right) ^{-\alpha }\,_2F_1\left( \alpha
,\alpha +1-\gamma ,\alpha +1-\beta ;\frac 1z
\right) \\
& + &\frac{\Gamma \left( \gamma \right) \Gamma \left( \alpha
-\beta
\right) }{
\Gamma \left( \alpha \right) \Gamma \left( \gamma -\beta
\right) }\left( -z\right) ^{-\beta }\,_2F_1\left( \beta ,\beta
+1-\gamma ,\beta +1-\alpha ;\frac 1z\right) ,\left| z\right|
>1.  \nonumber
\end{eqnarray}
We then obtain:
\begin{eqnarray*}
B\left( z,a,b\right)  &=&a^{-1}z^a\,_2F_1\left(a,1-b,a+1;z
\right)\\
& = &a^{-1}z^a\left\{ \frac{\Gamma \left( a+1\right) \Gamma
\left( 1-a-b\right) }{\Gamma \left( 1-b\right) \Gamma \left(
1\right) }\left(
-z\right) ^{-a}\,_2F_1\left( a,0,a+b;\frac 1z\right) \right.+
\\
&&\left. \frac{\Gamma \left( a+1\right) \Gamma \left(
a+b-1\right) }{\Gamma \left( a\right) \Gamma \left( a+b\right)
}\left( -z\right)
^{b-1}\,_2F_1\left( 1-b,1-a-b,2-a-b;\frac 1z\right) \right\}
\\
&=&\frac{\Gamma \left( a+1\right) }a\left\{ \left( -1\right)
^{-a} \frac{\Gamma \left( 1-a-b\right) }{\Gamma \left(
1-b\right) }+ \right.
\\
&&\left. \left( -1\right) ^{b-1}\,\frac{\Gamma \left(
a+b-1\right) }{\Gamma \left( a\right) \Gamma \left( a+b\right)
}\left( \frac 1z\right)
^{1-a-b}\,_2F_1\left( 1-a-b,1-b,2-a-b;\frac 1z\right) \right\}
\\
&=&\left( -1\right) ^{-a}\frac{\Gamma \left( a+1\right) \Gamma
\left( 1-a-b\right) }{a\Gamma \left( 1-b\right) }+\left(
-1\right) ^bB\left( \frac 1z,1-a-b,b\right)
\end{eqnarray*}
and therefore
\begin{equation}
G\left( k,z_1,z_2\right) =-\frac 12\left\{ B\left( \frac
1{z_2^2},-\frac k2,\frac 12\right) -B\left( \frac
1{z_1^2},-\frac k2,\frac 12\right) \right\} ,\;k\neq 2m\;\left(
m=0,1,2,\ldots \right)
\end{equation}
which can be computed using a continued fraction method
\citep[see][]{NRC92}. For situations when $k$ is close to
$2m\;\left( m=0,1,2,\ldots \right) $ we can use the
relationship
\begin{equation}
B\left( z,a,b\right) =B\left( a,b\right) -B\left( 1-z,b,a
\right)
\end{equation}
to get:
\begin{equation}
G\left( k,z_1,z_2\right) =\frac 12\left\{ B\left( 1-\frac
1{z_2^2},\frac 12,-\frac k2\right) -B\left( 1-\frac 1{z_1^2},
\frac 12,-\frac k2\right) \right\} .
\end{equation}
Finally,
\begin{equation}\label{eqn:A2PDelta}
P_\delta \left( \delta \right) =\frac C{2\delta ^2}\left\{
B\left( 1-\frac 1{\gamma _2^2},\frac 12,-\frac k2\right)
-B\left( 1-\frac 1{f^2\left( \delta \right) },\frac 12,-\frac
k2\right) \right\}.
\end{equation}

\begin{deluxetable}{clrcc}
\tablecolumns{5}
\tabletypesize{\footnotesize}
\tablewidth{0pt}
\tablecaption{\label{tab:Sample} The MOJAVE sample.}
\tablehead{ \colhead{B1950 Name} & \colhead{z} &
\colhead{$S_{15}$, Jy} & \colhead{FR class} &
\colhead{Optical class} \\
\colhead{(1)} & \colhead{(2)} & \colhead{(3)} & \colhead{(4)} &
\colhead{(5)}} \startdata
\cutinhead{MOJAVE sources of FR~II class}
0007$+$106 & $0.0893$ & $2.300$ & $2$ & G \\
0016$+$731 & $1.781$ & $2.260$ & $2$\tablenotemark{\dagger} & Q \\
0048$-$097 & \nodata & $2.160$ & $2$ & BL \\
0059$+$581 & $0.643$ & $3.307$ & $2$\tablenotemark{\dagger} & Q \\
0106$+$013 & $2.107$ & $2.972$ & $2$ & Q \\
0133$+$476 & $0.859$ & $4.953$ & $2$\tablenotemark{\dagger} & Q \\
0202$+$149 & $0.405$ & $2.293$ & $2$\tablenotemark{\dagger} & Q \\
0202$+$319 & $1.466$ & $2.284$ & $2$ & Q \\
0212$+$735 & $2.367$ & $2.842$ & $2$\tablenotemark{\dagger} & Q \\
0215$+$015 & $1.715$ & $1.532$ & $2$ & BL/HPQ \\
0224$+$671 & $0.523$ & $2.450$ & $2$ & Q \\
0234$+$285 & $1.207$ & $4.045$ & $2$ & Q \\
0235$+$164 & $0.940$ & $1.731$ & $2$ & BL/HPQ \\
0333$+$321 & $1.263$ & $2.249$ & $2$ & Q \\
0336$-$019 & $0.852$ & $3.452$ & $2$ & Q \\
0403$-$132 & $0.571$ & $2.750$ & $2\dagger$ & Q \\
0415$+$379 & $0.0491$ & $5.976$ & $2$ & G \\
0420$-$014 & $0.915$ & $10.438$ & $2$ & Q \\
0458$-$020 & $2.291$ & $2.325$ & $2$\tablenotemark{\dagger} & Q \\
0528$+$134 & $2.070$ & $7.945$ & $2$\tablenotemark{\dagger} & Q \\
0529$+$075 & $1.254$ & $1.630$ & $2$ & Q \\
0529$+$483 & $1.162$ & $1.662$ & $2$ & Q \\
0552$+$398 & $2.363$ & $5.020$ & $2$\tablenotemark{\dagger} & Q \\
0605$-$085 & $0.872$ & $2.797$ & $2$ & Q \\
0607$-$157 & $0.324$ & $7.263$ & $2$\tablenotemark{\dagger} & Q \\
0642$+$449 & $3.408$ & $4.310$ & $2$\tablenotemark{\dagger} & Q \\
0727$-$115 & $1.591$ & $5.125$ & $2$\tablenotemark{\ddagger} & Q \\
0730$+$504 & $0.720$ & $1.440$ & $2$ & Q \\
0736$+$017 & $0.191$ & $2.649$ & $2$ & Q \\
0738$+$313 & $0.630$ & $2.868$ & $2$ & Q \\
0748$+$126 & $0.889$ & $3.248$ & $2$ & Q \\
0804$+$499 & $1.432$ & $2.380$ & $2$ & Q \\
0805$-$077 & $1.837$ & $3.488$ & $2$\tablenotemark{\ddagger} & Q \\
0814$+$425 & $0.245$ & $1.810$ & $2$ & BL/HPQ \\
0827$+$243 & $0.941$ & $1.989$ & $2$ & Q \\
0836$+$710 & $2.218$ & $2.237$ & $2$ & Q \\
0906$+$015 & $1.018$ & $2.735$ & $2$ & Q \\
0917$+$624 & $1.446$ & $1.970$ & $2$\tablenotemark{\dagger} & Q \\
0923$+$392 & $0.698$ & $12.683$ & $2$ & Q \\
0945$+$408 & $1.252$ & $1.589$ & $2$ & Q \\
0955$+$476 & $1.873$ & $1.715$ & $2$ & Q \\
1038$+$064 & $1.265$ & $1.846$ & $2$\tablenotemark{\dagger} & Q \\
1045$-$188 & $0.595$ & $2.339$ & $2$\tablenotemark{\ddagger} & Q \\
1055$+$018 & $0.888$ & $5.296$ & $2$ & BL/HPQ \\
1124$-$186 & $1.048$ & $2.819$ & $2$\tablenotemark{\ddagger} & Q \\
1127$-$145 & $1.187$ & $3.388$ & $2$ & Q \\
1150$+$812 & $1.250$ & $1.651$ & $2$ & Q \\
1156$+$295 & $0.729$ & $3.302$ & $2$\tablenotemark{\dagger} & Q \\
1219$+$044 & $0.965$ & $1.678$ & $2$\tablenotemark{\ddagger} & Q \\
1222$+$216 & $0.435$ & $1.795$ & $2$ & Q \\
1226$+$023 & $0.158$ & $41.399$ & $2$ & Q \\
1253$-$055 & $0.538$ & $24.887$ & $2$ & Q \\
1308$+$326 & $0.997$ & $3.982$ & $2$ & BL/HPQ \\
1324$+$224 & $1.400$ & $1.953$ & $2$\tablenotemark{\dagger} & Q \\
1334$-$127 & $0.539$ & $8.868$ & $2$\tablenotemark{\dagger} & Q \\
1417$+$385 & $1.832$ & $1.772$ & $2$\tablenotemark{\dagger} & Q \\
1458$+$718 & $0.904$ & $2.740$ & $2$ & Q \\
1502$+$106 & $1.839$ & $1.956$ & $2$ & Q \\
1504$-$166 & $0.876$ & $2.031$ & $2$\tablenotemark{\ddagger} & Q \\
1510$-$089 & $0.360$ & $2.939$ & $2$ & Q \\
1546$+$027 & $0.412$ & $2.833$ & $2$ & Q \\
1548$+$056 & $1.422$ & $2.917$ & $2$\tablenotemark{\dagger} & Q \\
1606$+$106 & $1.226$ & $2.306$ & $2$ & Q \\
1611$+$343 & $1.401$ & $5.672$ & $2$ & Q \\
1633$+$382 & $1.807$ & $4.289$ & $2$ & Q \\
1637$+$574 & $0.751$ & $1.875$ & $2$\tablenotemark{\dagger} & Q \\
1638$+$398 & $1.666$ & $1.608$ & $2$ & Q \\
1641$+$399 & $0.594$ & $8.730$ & $2$ & Q \\
1655$+$077 & $0.621$ & $2.091$ & $2$ & Q \\
1726$+$455 & $0.714$ & $2.184$ & $2$ & Q \\
1730$-$130 & $0.902$ & $10.967$ & $2$ & Q \\
1739$+$522 & $1.379$ & $1.766$ & $2$ & Q \\
1741$-$038 & $1.057$ & $7.012$ & $2$\tablenotemark{\dagger} & Q \\
1758$+$388 & $2.092$ & $1.745$ & $2$\tablenotemark{\dagger} & Q \\
1800$+$440 & $0.663$ & $1.476$ & $2$ & Q \\
1803$+$784 & $0.680$ & $2.543$ & $2$ & BL/HPQ \\
1823$+$568 & $0.664$ & $2.309$ & $2$ & BL/HPQ \\
1828$+$487 & $0.692$ & $2.010$ & $2$ & Q \\
1849$+$670 & $0.657$ & $1.708$ & $2$ & Q \\
1928$+$738 & $0.303$ & $3.833$ & $2$ & Q \\
1936$-$155 & $1.657$ & $2.439$ & $2$\tablenotemark{\dagger} & Q \\
1957$+$405 & $0.0561$ & $1.680$ & $2$ & G \\
1958$-$179 & $0.652$ & $2.670$ & $2$\tablenotemark{\dagger} & Q \\
2005$+$403 & $1.736$ & $2.767$ & $2$\tablenotemark{\ddagger} & Q \\
2008$-$159 & $1.180$ & $2.134$ & $2$\tablenotemark{\ddagger} & Q \\
2037$+$511 & $1.687$ & $2.337$ & $2$ & Q \\
2121$+$053 & $1.941$ & $3.744$ & $2$\tablenotemark{\dagger} & Q \\
2128$-$123 & $0.501$ & $3.182$ & $2$\tablenotemark{\ddagger} & Q \\
2131$-$021 & $1.285$ & $2.439$ & $2$\tablenotemark{\dagger} & BL/HPQ \\
2134$+$004 & $1.932$ & $6.336$ & $2$ & Q \\
2136$+$141 & $2.427$ & $2.75$ & $2$\tablenotemark{\ddagger} & Q \\
2145$+$067 & $0.999$ & $10.372$ & $2$\tablenotemark{\dagger} & Q \\
2201$+$171 & $1.076$ & $1.986$ & $2$ & Q \\
2201$+$315 & $0.298$ & $3.27757$ & $2$ & Q \\
2209$+$236 & $1.125$ & $1.620$ & $2$\tablenotemark{\dagger} & Q \\
2216$-$038 & $0.901$ & $2.536$ & $2$ & Q \\
2227$-$088 & $1.562$ & $2.150$ & $2$ & Q \\
2230$+$114 & $1.037$ & $4.855$ & $2$\tablenotemark{\dagger} & Q \\
2243$-$123 & $0.630$ & $2.559$ & $2$\tablenotemark{\dagger} & Q \\
2251$+$158 & $0.859$ & $12.084$ & $2$ & Q \\
2331$+$073 & $0.401$ & $1.552$ & $2$ & Q \\
2345$-$167 & $0.576$ & $2.536$ & $2$\tablenotemark{\dagger} & Q \\
2351$+$456 & $1.986$ & $1.814$ & $2$\tablenotemark{\dagger} & Q \\
\cutinhead{MOJAVE sources of uncertain FR class}
0003$-$066 & $0.347$ & $3.302$ & \nodata & BL \\
0109$+$224 & \nodata & $1.654$ & \nodata & BL \\
0119$+$115 & $0.570$ & $2.007$ & \nodata & BL/HPQ \\
0300$+$470 & \nodata & $1.770$ & \nodata & BL \\
0422$+$004 & \nodata & $1.739$ & \nodata & BL \\
0446$+$112 & \nodata & $2.256$ & \nodata & U \\
0648$-$165 & \nodata & $3.437$ & \nodata & U \\
0716$+$714 & $>0.52$ & $2.586$ & \nodata & BL \\
0735$+$178 & $>0.424$ & $1.635$ & \nodata & BL \\
0754$+$100 & $0.266$ & $1.833$ & \nodata & BL/HPQ \\
0808$+$019 & $1.148$ & $1.590$ & \nodata & BL \\
0823$+$033 & $0.506$ & $2.467$ & \nodata & BL/HPQ \\
0829$+$046 & $0.180$ & $1.720$ & \nodata & BL \\
0851$+$202 & $0.306$ & $4.375$ & \nodata & BL/HPQ \\
1036$+$054 & \nodata & $2.664$ & \nodata & U \\
1213$-$172 & \nodata & $2.564$ & \nodata & U \\
1413$+$135 & $0.247$ & $1.719$ & \nodata & BL \\
1538$+$149 & $0.605$ & $1.630$ & \nodata & BL/HPQ \\
1749$+$096 & $0.320$ & $6.020$ & \nodata & BL/HPQ \\
1751$+$288 & \nodata & $2.015$ & \nodata & U \\
2021$+$317 & \nodata & $2.158$ & \nodata & U \\
2155$-$152 & $0.672$ & $2.147$ & \nodata & BL/HPQ \\
2200$+$420 & $0.0686$ & $5.669$ & \nodata & BL/HPQ \\
2223$-$052 & $1.404$ & $6.572$ & \nodata & BL/HPQ \\
\cutinhead{Excluded MOJAVE sources}
0238$-$084 & $0.0049$ & $2.481$ & $1$ & G \\
0316$+$413 & $0.01756$ & $12.908$ & $1$ & G \\
0430$+$052 & $0.033$ & $4.412$ & $1$ & G \\
0742$+$103 & $2.624$ & $1.504$ & \nodata & Q \\
1228$+$126 & $0.00436$ & $2.969$ & $1$ & G \\
2021$+$614 & $0.227$ & $2.735$ & \nodata & G \\
\enddata
\tablecomments{\small Column (1): Source B1950 name;
Column (2): redshift; Column (3): Flux density at $15\GHz$ in
Jy; Column (4): Fanaroff-Riley class; Column (5): Optical
class: BL=BL~Lac, Q=quasar, G=radio galaxy, BL/HPQ=BL~Lac/High
Polarization Quasar \citep[e.g.,][]{Veron00}, U=unidentified.}
\tablenotetext{\dagger}{Classification based on luminosity at
$1.4\GHz$ when the morphology was inconclusive.}
\tablenotetext{\ddagger}{Classification based on luminosity at
$1.4\GHz$ when a kpc-scale image was unavailable.}
\end{deluxetable}

\begin{deluxetable}{cllc}
\tablecolumns{4}
\tabletypesize{\footnotesize}
\tablewidth{0pt}
\tablecaption{\label{tab:BestFitParams}Best fit model
luminosity function parameters.
%($L_*=10^{27}\WHz$)
} \tablehead{\colhead{Parameter} & \colhead{FR~II only} &
\colhead{All except FR~I} & \colhead{Units}} \startdata
  $\alpha$ &  $-2.53 \pm 0.06$ & $-2.65 \pm 0.06$ \\
  $m$ &       $1.4 \pm 0.1$ & $1.6 \pm 0.1$ \\
  $z_0$ &     $1.29 \pm 0.09$ & $1.18 \pm 0.09$ \\
  $\sigma$ &  $0.76 \pm 0.09$ & $0.8 \pm 0.1$ \\
  $n_0$ & $(2.87 \pm 0.04) \times 10^{-10}$ & $(2.22 \pm
  0.04)\times 10^{-10}$ & $\MpcCubed$ \\
  $\rho$ &  $(1.579 \pm 0.008) \times 10^3$ & $(4.39 \pm
  0.07) \times 10^3$ & $\GpcCubed$ \\
  $\mbox{K}$ &             $(5.49 \pm 0.04)\times 10^9$ &
  $(1.55 \pm 0.03)\times 10^{11}$ \\
  $\mathcal{L}_1$ & $10^{22.2}$ & $10^{21.6}$ & $\WHz$ \\
  $\mathcal{L}_2$ & $10^{29.1}$ & $10^{29.2}$ & $\WHz$ \\
\enddata
\tablecomments{\small The errors in the parameters
$\alpha, m, z_0$, and $\sigma$ of the model LF were computed
using the $\Delta S=1$ method as described in the text. The
error estimates on the normalization factor $n_0$, space
density $\rho$ for $L>1.3\times 10^{25}\WHz$ and parent
population $K$ were calculated using their cumulative
distribution functions as described in subsection
\ref{subsec:ModelPar}.}
\end{deluxetable}

\end{document}